# SharedCanvas: A Collaborative Model for Medieval Manuscript Layout Dissemination

Robert Sanderson
Los Alamos National Laboratory
Los Alamos
NM 87544, USA
+1 (505) 665-5804
rsanderson@lanl.gov

Benjamin Albritton
Stanford University
Stanford
CA 94305, USA
+1 (605) 387-8678
blalbrit@stanford.edu

Rafael Schwemmer
e-codices
Rue de l'Hopital 4
CH-1700 Fribourg, Switzerland
+41 (26) 300-7919
rafael.schwemmer@unifr.ch

Herbert Van de Sompel
Los Alamos National Laboratory
Los Alamos
NM 87544, USA
+1 (505) 667-1267
herbertv@lanl.gov

## ABSTRACT

In this paper we present a model based on the principles of Linked Data that can be used to describe the interrelationships of images, texts and other resources to facilitate the interoperability of repositories of medieval manuscripts or other culturally important handwritten documents. The model is designed from a set of requirements derived from the real world use cases of some of the largest digitized medieval content holders, and instantiations of the model are intended as the input to collection-independent page turning and scholarly presentation interfaces. A canvas painting paradigm, such as in PDF and SVG, was selected based on the lack of a one to one correlation between image and page, and to fulfill complex requirements such as when the full text of a page is known, but only fragments of the physical object remain. The model is implemented using technologies such as OAI-ORE Aggregations and OAC Annotations, as the fundamental building blocks of emerging Linked Digital Libraries. The model and implementation are evaluated through prototypes of both content providing and consuming applications. Although the system was designed from requirements drawn from the medieval manuscript domain, it is applicable to any layout-oriented presentation of images of text.

## 1. INTRODUCTION

There are many repositories and Digital Libraries (DLs) that maintain digitized page images of medieval manuscripts or other historically important, handwritten documents. These images are often the only way in which scholars and students can interact with the material. Use of the digital surrogate increases the likelihood of its persistence, and interactions with the physical copy decrease its usable lifetime, making the use of surrogates attractive to the owning institution as well as to humanities travel budget managers. Therefore, it is essential that the digital surrogate be as rich an experience as possible for the scholar, with access to the existing scholarship about the manuscript, as discussed by Audenaert and Furuta [1]. The surrogate for the physical object is the humanist's primary research data.

Institutions holding medieval manuscripts have long known the value of digitization and millions of grant and industry dollars have been spent generating images of decaying physical pages, yet less than 1% of existing medieval documents have been digitized to date. While the digitized manuscripts are unique, much of the effort to display the digitized material has been duplicated across institutions with each recreating very similar basic page-turning applications; the only differences being for institutional branding and the seemingly unique complexities of their documents. Further, practically all of the presentation effort has been used for navigation within the institutional silos of images and texts, rather than cross-collection capabilities.

At a series of meetings[1] of content providers, scholars and technologists from organizations such as the British Library, the National Library of France, Google Freebase, the University of Oxford, and the authors' institutions, it was recognized that in order to reduce the duplication of effort, a single shared model for description of manuscripts was necessary. Such a model could be instantiated for various digital manuscript collections and provided to a conforming display application. Furthermore, the "unique" rendering complexities of many institutions' documents were, in fact, shared to a very large degree and hence such a display application would require minimal, if any, customization beyond branding. If this shared data model was at the same time sufficiently simple to allow for comprehension and implementation, and sufficiently expressive to facilitate the description of the multi-structured documents, then access to the digital surrogates would be greatly improved and the duplicated time and effort could be rededicated to further digitization or more complete descriptions and transcriptions.

[1] http://dmstech.group.stanford.edu/

A second goal identified at the meetings was to extend the notion of interoperability between manuscript repositories from sharing of their resources to seamless integration between them. While displaying an image is a common baseline capability, the experience of the scholar can be greatly enriched by involving materials from multiple repositories. Other repositories may contain the transcription of the text, processing services may be able to discover locations within the image of the transcribed text, and it should be possible to integrate this data into the display. Further value would be added by links to DLs containing publications regarding the manuscript or related material, scientific data about the subject matter of the manuscript, or scholarly annotations. Such capabilities would yield a coherent landscape of interconnected systems, rather than the current set of disparate content silos.

This paper describes the steps taken towards meeting these goals and, in Section 2, details the use cases and requirements that were generated from the medieval manuscript domain. The background work is described in Section 3, and the abstract SharedCanvas model in Section 4. The technologies used to instantiate the model are discussed in Section 5, and experiments evaluating both the expressiveness for describing manuscripts and the investment required for implementation are described in Section 6.

## 2. REQUIREMENTS AND USE CASES

The goal of this research is to provide a standardized description of the digital resources that are surrogates for culturally important, primarily textual physical objects in order to enable interoperability between repositories, tools, services and presentation systems. The primary domain of use is medieval manuscripts that have had their pages individually digitized and their text transcribed; the physical Item is the artifact of interest, rather than the Work in the FRBR sense. Several basic requirements were derived from this domain and then expanded upon via the examination of more complex use cases, which were beyond the capabilities of existing systems. These extended requirements inform the design of the model, and therefore will be discussed in detail.

Requirements were identified in four main areas:

1. **Images** and their relationships with the physical object
2. **Texts** and their relationships with the images
3. **Sequencing** of the Images and Texts
4. **Rendering** of the Images and Texts

### 2.1 Image Requirements

While in the most basic case there is exactly one image per manuscript page, multiple images will often exist. Those images may differ in color, size, depth, lighting conditions or scale due to the positioning of either the page or the camera during digitization, and rectification through dewarping transformations should be possible, as discussed by Baumann and Seales [4]. Hence, the model must contain sufficient information to determine the most appropriate image for the user. In order to avoid unnecessary duplication of information, the model must also provide a means to avoid attaching information directly to any single image when it also applies to other images.

Images may also exist that depict only parts of the manuscript page, such as very high-resolution images of beautiful illuminations or decorated initial letters. In these cases, the entire page may be left undigitized or only available at a much lower resolution. This is especially true of older digitization projects where only microfilms of the non-illuminated pages are available. From a modeling perspective, the ability to connect the high quality image with the appropriate part of the page in the full image is important.

The converse is also true, where part of the image depicts the entire page. Depending on the digitization workflow, the image may depict more than just the page, such as a border around the outside of the scanning bed or calibration tools such as color strips or rulers, which would not be desirable to show to all users. In this case it is necessary for the model to be able to describe the area within the image that depicts only the page.

A single image may depict multiple pages, such as an image of an open spread of two pages. If these two pages have pictures or text that cross between them (as in Figure 1), it is important to have mappings from the parts of the image to the page, and a method for describing the spread as a whole. Again, the model must avoid the duplication of information between the spread and the individual pages.

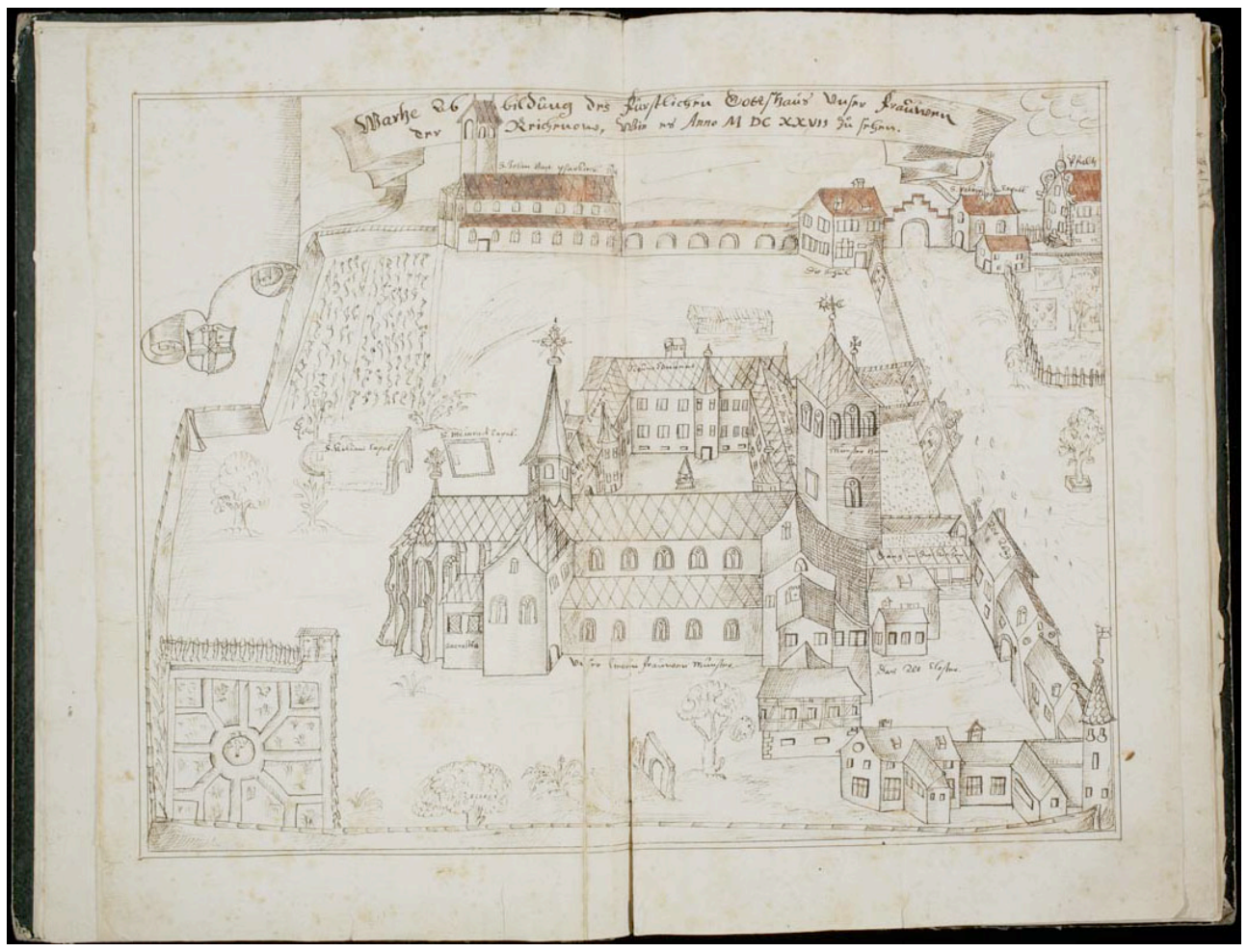

**Figure 1. Open Spread of Two Pages; Y112 [22]**

There are many cases in which only fragments (parts of the original page) remain. As an example, Figure 2 depicts a manuscript from the Abbey Library of St. Gall, where two fragments, likely not from the same original page, have been bound together. The volume collects together fragments from the 4th through 15th centuries, and was assembled in 1822 [15]. When a fragment is digitized separately, the image depicts only part of the original page. A single image may also depict multiple small fragments, regardless of where they were originally located. The fragments might be housed together in the same container, or stuck to a further page or glass slide. In order to display the fragment in conjunction with other fragments from the same page, individual parts of the image must be able to be mapped in the model to the appropriate locations. Fragments of pages are typically irregularly shaped and hence the model must also be able to make use of arbitrary polygons rather than just rectangular bounding boxes to describe the parts of the images.

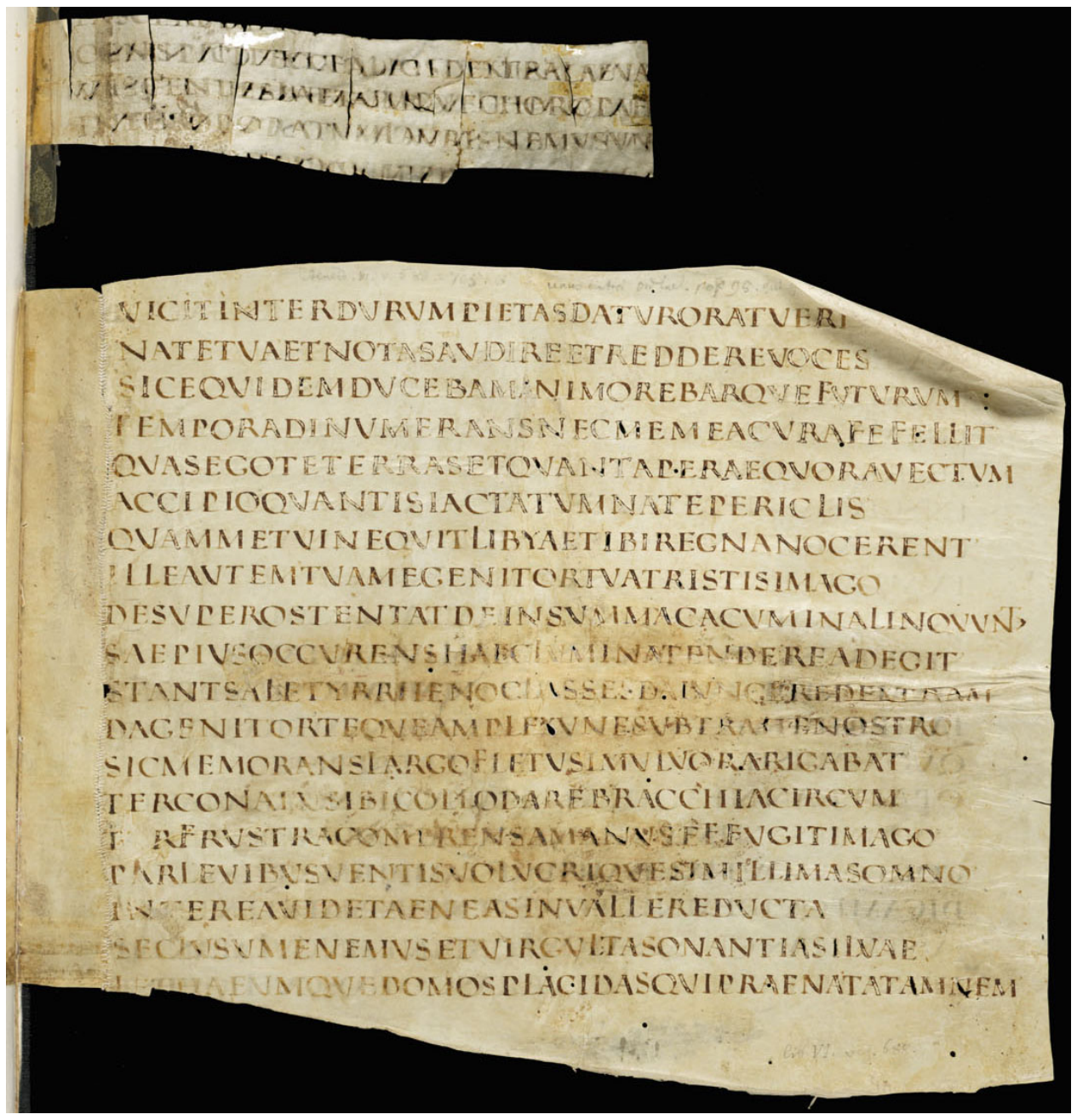

**Figure 2. Two Fragments; Cod. Sang. 1394 p. 31 [15]**

Equally, there may not be any image that depicts a page at all. This might be because the digitization process would irrevocably damage the page, or it may no longer exist but either is known or hypothesized to have existed in the past. Despite the lack of an image, the model must still be able to account for the existence of the page and allow other information to be associated with it.

## 2.2 Text Requirements

The semantic properties of text have long been understood and appropriate markup languages exist to describe these features, so this is not a direct research concern. Instead the focus is on the relationships between text and image.

All of the relationships between image and the physical resource depicted exist in parallel for the transcribed text and its depiction. The text may exist in multiple copies; it may be a complete resource or part of a larger resource such as a TEI XML file; linking may be possible at different levels of granularity to the appropriate part of the image; the areas in the image depicting the text are likely to be non-rectangular; and so forth.

In the case where an image of the page is not available, the text may still be known from other copies, and this information should be available to the reader without a depiction.

The most challenging textual use cases are palimpsests, manuscripts where one or more texts have been erased, and the pages reused for another text. The original texts can be recovered using techniques such as multispectral imaging as in the case of the Archimedes Palimpsest, fully described by Reviel Netz and William Noel [23]. Some images may therefore depict two or more completely separate texts, often perpendicular to each other. The identity of each text is important, and could require rotation of either the image or rendered text for verisimilitude. The model must record this information and not assume a one to one relationship between page or image and text, nor any mandatory rotational alignment. If the manuscript leaves were rebound when the second text was written, the first text would need a different page order. This brings us to page and text ordering requirements.

## 2.3 Sequencing Requirements

Most page turning applications assume a single correct order for the pages, yet many manuscripts have been disassembled and rebound over the centuries by well-meaning curators, and intentionally or not, the page order has changed over time. It should be possible to reconstruct the order as it was at a particular point in time, without duplicating all of the resources.

The presentation of alternate paths within the same order is also an important use case when, as per Figure 1, a spread exists as a single image but also as separate pages. An animated page turning application, for example, should not try to display the image of the spread as a single page, nor should it animate the turn using the entire image. Yet for scholars, access to the image of the full spread is important to get as accurate a depiction as possible.

The description of subsets of pages from the ordered list is also important. In the manuscript construction process pages are collected together in quires (sets of normally 16 pages), and the boundaries between these is important information to scholars. Textual sections such as chapters or verses are equally important for the humanist interested in particular parts of the text. Other features, such as the range of pages at the beginning or end of a manuscript that do not contain any text, are also important for navigation and display of an automatically generated table of contents. In order to enable the description of sections at any granularity, these sub-lists must be able to include parts of a page, since textual and other features do not necessarily align with page breaks.

When the text is transcribed line by line, it is important to be able to explicitly describe the reading order of those lines. Visual clues, such as size, location and color of the writing, may make the order clear for human readers, but are difficult to interpret for a machine due to writing in the margins (marginalia), writing between regular lines (interstitial text), decorated initials and the scribal tricks used to justify text into columns. The ability to express the correct sequence of textual resources is just as important as the order in which to display the pages.

## 2.4 Rendering Requirements

Many of the rendering requirements have already been discussed with respect to the Images, Texts and Sequencing. Further requirements include the ability to create and display scholarly annotations that discuss the images and text at a very fine level of granularity.

The creation and maintenance of the manuscript description should be possible in a highly distributed and collaborative environment, to enable the sharing of content and expertise between different communities and individuals. The rendering application must be prepared to consume and display resources from across many locations, not limited to a single file or content silo as is the case currently.

Visualizing the resources and the relationships between them in an innovative way that promotes scholarship is an interesting research challenge for future work.

# 3. BACKGROUND RESEARCH

Research was needed to discover if any existing systems or models met the above requirements. While much related work has been done, interoperability and the complexities of medieval manuscripts have not been primary research topics. Again we discuss in terms of the four areas of concern.

## 3.1 Image Layout Analysis

Baechler et al.'s recent work on a layout model [2] for manuscripts looks promising in its acknowledgement of the complexities of the issues and its demonstration of the need for non-rectangular bounding areas for line segmentation. Unfortunately their model is designed for evaluating automated methods of segmentation, and not as an interoperability mechanism. Their four layers of text, physical medium, illuminations and commentary would be insufficient for describing palimpsests, and their XML document structure inappropriate for ease of cross-institution collaboration given the expectation with XML of a self-contained document.

The analysis of the archive of the Dutch queen [9], which focuses on extracting information from a handwritten table of contents document, is also focused on layout. It also demonstrates complexities such as arbitrarily shaped boundaries, but is not a general solution.

## 3.2 Text and Image Linking

The work of Brugman et al. [8] provides a strong baseline model for linking resources describing cultural heritage objects through the use of annotations; their first example is transcription of an image with additional semantics linked to the text. In their description of ARMARIUS [12], Doumat and colleagues give a model for web-based annotation of digitized medieval manuscripts. However, in both cases the annotations are modeled as graph relationships between individual images and texts, and there is no notion of a method to transcribe without an image, of transferability of information between equivalent images or of services providing scaled and tiled images.

Other modeling focuses only on the textual requirements and ignores the relationship with images. Schmidt and Colomb [27] looks at models for online text with multiple versions and overlapping hierarchies, and Rehbein [24] considers the problem from the change of the text over time, as a medieval codex of law is updated over many years as the laws of the town were revised.

## 3.3 Sequencing

The modeling of multiple page orders is one of the aspects tackled by Bauer and Hernath [3] along with the use of offset annotations to describe differences of opinion about contentious transcriptions. Their system, tested with the previously mentioned Archimedes Palimpsest, has an Ordering and Indexing Layer separate from the text and images to enable multiple sequences. Their system does not have any image requirements, and overlays the annotations on top of the standard, tree-structured TEI [28].

## 3.4 Rendering

We do not consider the exact user interface or the details of simulating interaction with a physical object in this research. Marshall [20] and Liesaputra [18] have studied the page turning experience extensively and we defer to their knowledge.

Beyond the area of manuscripts, there are many well-known layout oriented systems capable of rendering images and text. The most well known is, of course, PDF [16] where images and text can be laid out on a blank, page sized canvas. The canvas notion is also used in SVG [11] (an XML description format primarily for vector graphics) and HTML5's canvas element [14] that can be drawn on using javascript functions.

Beyond the domain of documents altogether, the user interfaces of software applications are also often built up on empty canvases by adding layout boxes, controls, text and images. Mozilla's Firefox browser's user interface is constructed from a series of XML files describing the layout[2]. Programming libraries such as GTK[3] function in essentially the same way, but with function invocations rather than XML elements.

# 4. SHAREDCANVAS MODEL

At the foundation of the SharedCanvas model is the insight that there is not a one to one correlation between manuscript page and the depicting image. The common approach of linking a transcription or annotation to a point on a particular image is not appropriate, as convincingly illustrated by the case in which an image for the page does not exist. Also, in cases where multiple images do exist, such linking would need to be repeated for each of the images, some of which may not be known by the system making the connection. For fragments, it must be possible to create annotations that reference both digitized and undigitized sections of the manuscript.

## 4.1 Image and Text Layout

Following the lead from canvas based layout systems, the SharedCanvas model starts with a blank **Canvas** to be drawn on, which stands for a page in the manuscript. In Figure 3 the Canvas is depicted as a rectangle with a dashed black border. The Canvas has its own dimensions or aspect ratio that may or may not be the same as any image, however the top left hand corner of the canvas is chosen to correspond to the top left hand corner of the manuscript page, and similarly for the bottom right hand corners of canvas and page, as shown by the dashed orange lines.

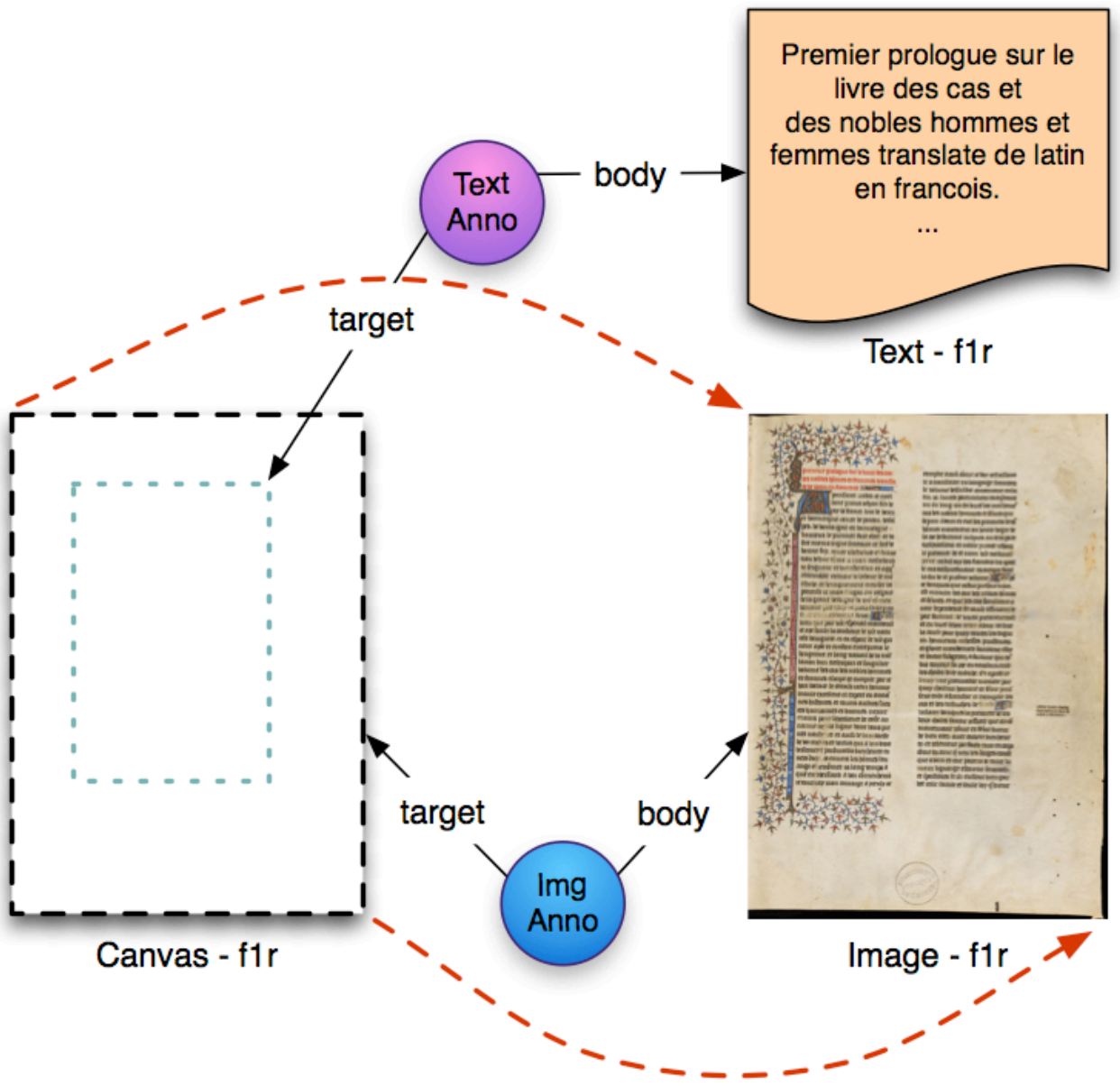

**Figure 3. Canvas, Image and Text; MS fr. 190/1 [7]**

Images and texts are then overlaid on top of the blank Canvas using an **Annotation Paradigm**, following the approach taken by Brugman, Bauer and others. By using annotations to paint the canvas, the number of technology dependencies is reduced, as the rendering of scholarly annotations is also required. The body of an **ImageAnnotation** is the Image and the target is the Canvas. In Figure 3, the "Img Anno" circle represents an annotation that associates the Image resource with the full Canvas.

[2] http://www.mozilla.org/projects/xul/

[3] http://www.gtk.org/

In order to paint either an image or text on the appropriate region of the Canvas, **Segment Information** is used. In the example, the information records the location of the bounding box depicted with a blue dotted line within the Canvas. The "Text Anno" node represents a **TextAnnotation** that associates the transcribed text with the region of the Canvas in which the text is located, potentially to a curve rather than straight line. The Image or Text itself may also have Segment Information if only part of a larger resource is needed, such as a section of a TEI transcription. A non-rectangular manuscript would use this approach to paint only to the appropriate segments of the Canvas.

The body of an annotation may also be a set of equivalent resources from which a choice is made by the presentation application. An **ImageChoice** set could then contain all the equivalent images of a page, to be applied to a Canvas, at different sizes, different color depths and so forth. The alignment of these multiple images is enabled by the independent coordinate system of the Canvas. The same set construction, a **TextChoice**, would allow for a choice of multiple Texts to be associated with a specific line segment. The application would then select the appropriate text based on properties such as author or the edition it is derived from.

Scholarly annotations can also be applied to any of the resources, as appropriate. If the annotation should be displayed regardless of the images and texts being used to render the Canvas, then it should target the Canvas. On the other hand, a criticism of a particular transcription should be attached to the transcription itself, so that if the transcription is replaced, the annotation would no longer be displayed. These annotations can come from any source and pull in resources from outside that otherwise would not have been known about.

The **Type** of annotation is used to convey the expected behavior to the presentation system. In Figure 3, the two different types of Annotation are Image Annotation and Text Annotation, however many more exist, such as different types of scholarly annotation. The different classes allow for the easy distinction between annotations that overlay images or transcriptions on the canvas and more traditional scholarly annotations. By recognizing the differences and similarities between these types of annotation, the display system should require less implementation effort than if the layout were done in a completely different method from regular annotations.

Areas called **Zones** may be delimited, annotated with Text and/or Images, and then painted onto multiple Canvases via **ZoneAnnotations**. This functionality facilitates many of the more challenging use cases, such as when pages should be displayed together as a spread, or in the palimpsest case when some images contain both texts and others only one. The Zones will maintain all of their associated annotations, permitting the display of the information without having to repeat it in multiple locations.

## 4.2 Sequencing

The annotated Canvas method allows us to build up the view of a single page piece by piece, and the order in which multiple Canvases, which make up the entire manuscript, should be presented to the user is recorded in a **Sequence**. In Figure 4, the Sequence is depicted as a light green circle labeled "Seq" at the top of the diagram. The Sequence is not necessarily a single, linear list as there may be alternative paths from one canvas to the next, such as either through two Canvases or via a single Canvas that represents the combined spread.

The same Canvas may appear in multiple Sequences to allow for rebinding or competing theories of provenance. Multiple Sequences might also be used in order to provide better navigation for different viewing platforms, such as to make the most appropriate use of a multi-column manuscript on a smart-phone's limited width display.

Other groupings of Canvases are also desirable in order to model textual or physical boundaries within the manuscript. These groupings are called **Ranges**, where all of the resources in the Range are Canvases or parts of Canvases from a single Sequence. A Range is depicted in Figure 4 as the darker green node labeled "Rng" at the bottom of the diagram. The example Range includes all of Canvas 3, and the lower part of Canvas 2. The reuse of the segmentation concept, depicted with the blue dotted line within Canvas 2, would enable the inclusion of the region in which the beginning of a new chapter is displayed, for example.

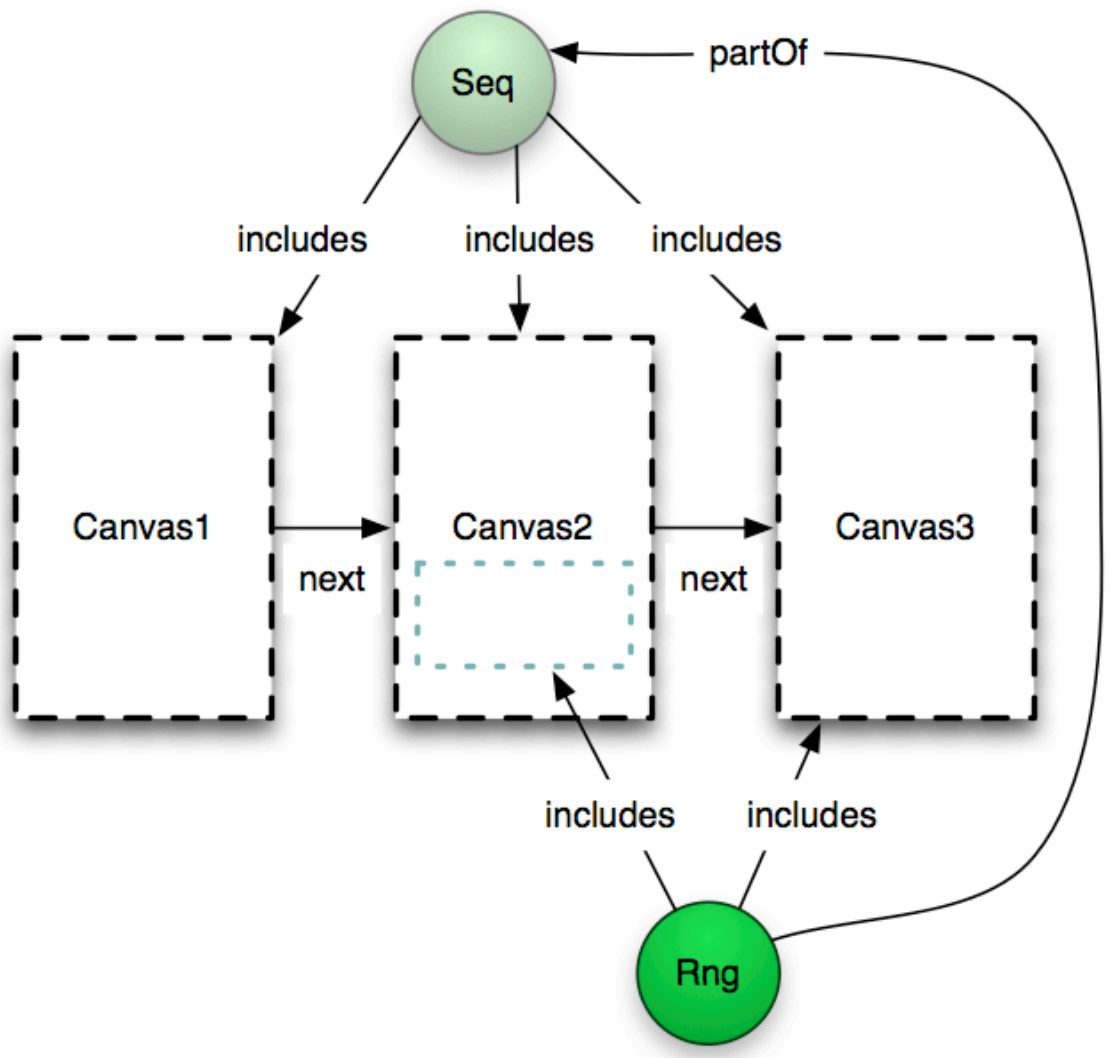

**Figure 4. Sequence and Range of Canvases**

As multiple Ranges may overlap, there is not a strict hierarchy of Sequence, then Range, then Canvas. Instead the Range must link to the Sequence of which it is part, and the Sequence should list all of the Ranges it knows about.

Groupings above the level of the Sequence are also important for discovery and presentation. In Figure 5, these groupings are the nodes depicted in the Discovery section above the dashed line. If the manuscript has been transcribed line by line, there may be many thousands of small annotations linking each line with the appropriate region of a Canvas. In order to satisfy the explicit reading order requirement, there must be an ordered collection at least per Canvas, if not across the entire Sequence. This is the "Text Ordr" node in Figure 5. For discovery purposes, collections of the ImageAnnotations would also be useful ("Img List"), as would the set of ZoneAnnotations ("Zone List"). Finally, as there may be multiple Sequences for a single manuscript, a single top-level **Manifest** is introduced that collects them together along with these sets of annotations.

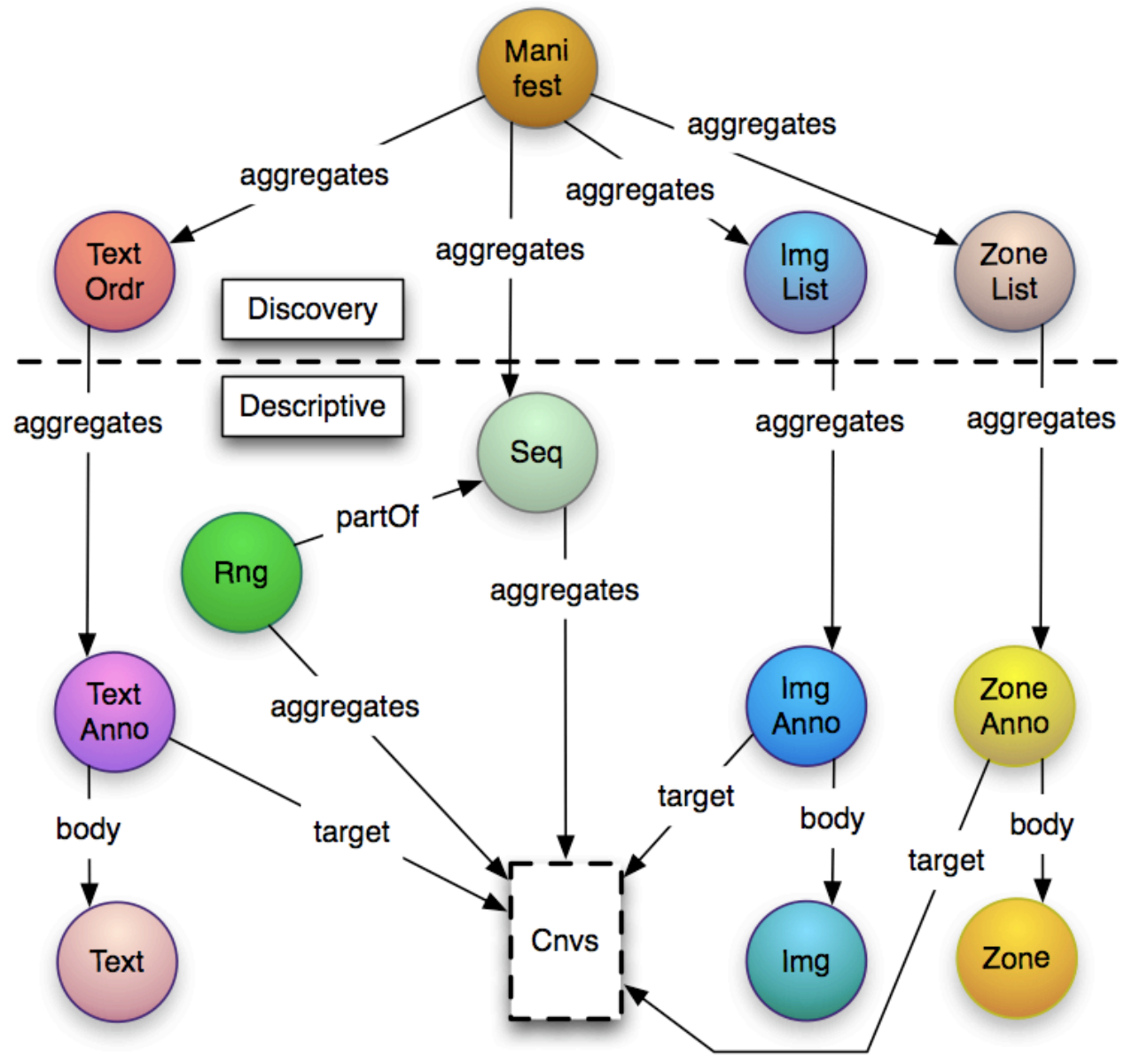

**Figure 5. Complete SharedCanvas Model**

# 5. INSTANTIATION

The SharedCanvas concepts were conceived to address the core functional requirements, but these need to be instantiated using specific technologies. Although most manuscript transcription and description work has taken place to date using XML, and notably with the TEI and ALTO schemas, many of the use cases require a graph rather than an XML tree. In order to permit the distributed creation and use of the model simultaneously by multiple tools and repositories, an approach that follows the architecture of the web [17] is necessary. Linked Open Data [6] using RDF [19] fit these requirements for a web-centric graph.

## 5.1 OAC Annotations for Layout

The main choice of ontology is for the Annotations used to overlay resources on the canvas and to permit scholars to annotate them. The Open Annotation Collaboration has defined an RDF based ontology [26], of which the Alpha3 version is current at the time of writing. OAC allows the annotation of any resource with any other resource irrespective of media type, unlike many of its predecessors in which the body is required to be textual. This feature is essential in order to overlay both images and transcriptions on the blank canvas using the same method. It is also important for the use case where multiple equivalent images are available, as the body is then a set of those images (see Section 5.2).

Furthermore, OAC has a graceful method of identifying selected parts of resources to be the target or the body of the annotation. This feature is crucial for the many use cases that require segmentation, such as manuscript fragments, removing the ruler or digitization housing from the presentation without modifying the images that may not be under the control of the presentation layer, and for presenting zones within canvases to ensure that data does not have to be added to or from multiple locations

The segmentation method put forward by OAC is twofold. First, if the segment can be described using the W3C's Media Fragment specification [29], then that approach is recommended. Media Fragments are a URI construction in which information describing the region of interest is embedded within the URI after a '#' character. For images this allows rectangular regions to be described, using the coordinates of the upper left corner, the height and width.

An example of the application of OAC where a rectangular section of an image, identified using a Media Fragment URI, should be overlaid on top of a Canvas, in the relatively readable Turtle [5] syntax:

```
:myAnno a oac:Annotation;
     oac:hasTarget :canvas1;
     oac:hasBody <image1#xywh=10,10,640,480>;
:canvas1 a sc:Canvas;
    exif:height 1024;
    exif:width 768;
```

As support for non-rectangular regions is also crucial, it is fortunate that OAC provides a second method for describing regions, called a Constraint. Constraints are separate resources that describe how to determine the region of interest in a media dependent fashion. For images, the recommended method is to use an SVG description, either embedded within the annotation document or referenced as an external resource. The same techniques used for images could easily be applied to Canvases, as they share the essential properties of height and width.

## 5.2 OAI-ORE Aggregations for Sequencing

The immediately obvious choice of ontology for a Sequence would be the use of an OAI-ORE Aggregation [31]. Aggregations are sets of resources plus some metadata, and the OAI-ORE ontology has been increasingly adopted by DL systems since its publication. However, the ubiquity of the ordering requirement in the use cases is troubling. As order is local to the Aggregation (the same resource may be in a different order in another Aggregation, an important concept given the rebinding use case), the ORE model prescribes the use of Proxy nodes that stand for the resource as it occurs in the Aggregation. The mandatory use of Proxies for every aggregated resource would almost double the number of nodes, and almost triple the number of relationships required for even the simplest scenario.

Thankfully, for this particular case, an alternative solution exists. RDF objects may have multiple classes, and thus a single resource may be an ORE Aggregation at the same time as it is an RDF List. The fundamental RDF List construction imposes an order through the use of many anonymous nodes, and due to its definition at the core of the standard, serializations have ways to avoid minting identifiers for all of these nodes. The anonymous nodes become part of the plumbing that is conveniently taken care of by RDF libraries, rather than something that needs to be handled by every application. Again in the Turtle syntax, a resource that is both an Aggregation and a List would be expressed as:

```
:mySequence a ore:Aggregation, rdf:List;
      ore:aggregates :page1, :page2, :page3,
                     ..., :pageN;
      rdf:first :page1;
      rdf:rest (:page2 :page3 ... :pageN );
```

In this way other systems that understand Aggregations can process the descriptions, further systems can easily process the List, and the thousands of Proxy URIs do not need to be minted.

A further advantage of this approach is that there may be aggregated resources that are not part of the order expressed in the List. This facility would allow for images of the edges, spine, container of the manuscript to be included, but not directly part of the page-turning sequence.

There is one case in which proxy nodes are still required: the alternate paths use case. In this case, there isn't a single linear List, and the Proxy construction from ORE comes to the rescue. From the requirements analysis, this use case is important but does not occur in the majority of manuscripts. Thus the use of ORE plus Lists provides an easy transition from the unordered set, through to simple lists, and on to more complex multi-pathways.

This construction can deal with orderings ranging from no order through to multiple pathways, the same approach can be used to model the other sets and lists needed by the model: Ranges, Manifests, the different Choices, and the Lists of Annotations.

# 6. EXPERIMENTATION

The majority of experimentation with the model has been to attempt to describe increasingly complex use cases. Initial prototype implementations were then created to produce the RDF serializations, and to in turn consume them and render the images and texts for a user. The implementations do not provide the full capabilities envisioned for a scholarly environment, only enough functionality to prove that the descriptions can be rendered at least as well as in their current interfaces.

## 6.1 Image and Text Layout

### 6.1.1 *Morgan Library Manuscript 804*

The first manuscript description generated was derived from Sanderson's Ph.D. work [25] on an electronic edition of Froissart's Chronicles. This description demonstrates most of the basic requirements throughout a full manuscript.

The manuscript described is held at the Morgan Library, with shelfmark M804 [13]. The electronic resources used consist of 515 low quality black and white images derived from a microfilm with each depicting a single side of a page, 10 color images depicting illuminated pages, the text marked up in TEI, and hand crafted locations of each line within each image.

The model generated for the first page is depicted in Figure 6. A Canvas was created for the page, and the appropriate black and white image associated with it via an Image Annotation. As the first page is richly illuminated, there is also a color image available, and thus the Annotation's body is an ImageChoice that includes both of these images. The size and color depth were recorded in properties that are not depicted.

The text was split up into lines and linked to the appropriate region of the Canvas via Text Annotations. Only the first four are depicted, and the different size and color represents aspects of the physical text recorded with additional properties. The location of the lines of text were transferred to the Canvas coordinates from those of the image with appropriate scaling. These became the rectangular bounding boxes used as segment information within the Canvas, and thus the targets of the Text Annotations were Media Fragment URIs that recorded x and y coordinates, plus height and width.

Other color images depicted only an illumination, and not the entire page. These detail images were treated as the body of Image Annotations that targeted the appropriate location within the Canvas using Media Fragment URIs, and could thus be overlaid above the lower quality digitized microfilm images.

A single Sequence was generated that ordered all of the 515 Canvases, the green 'M804' node in Figure 6.

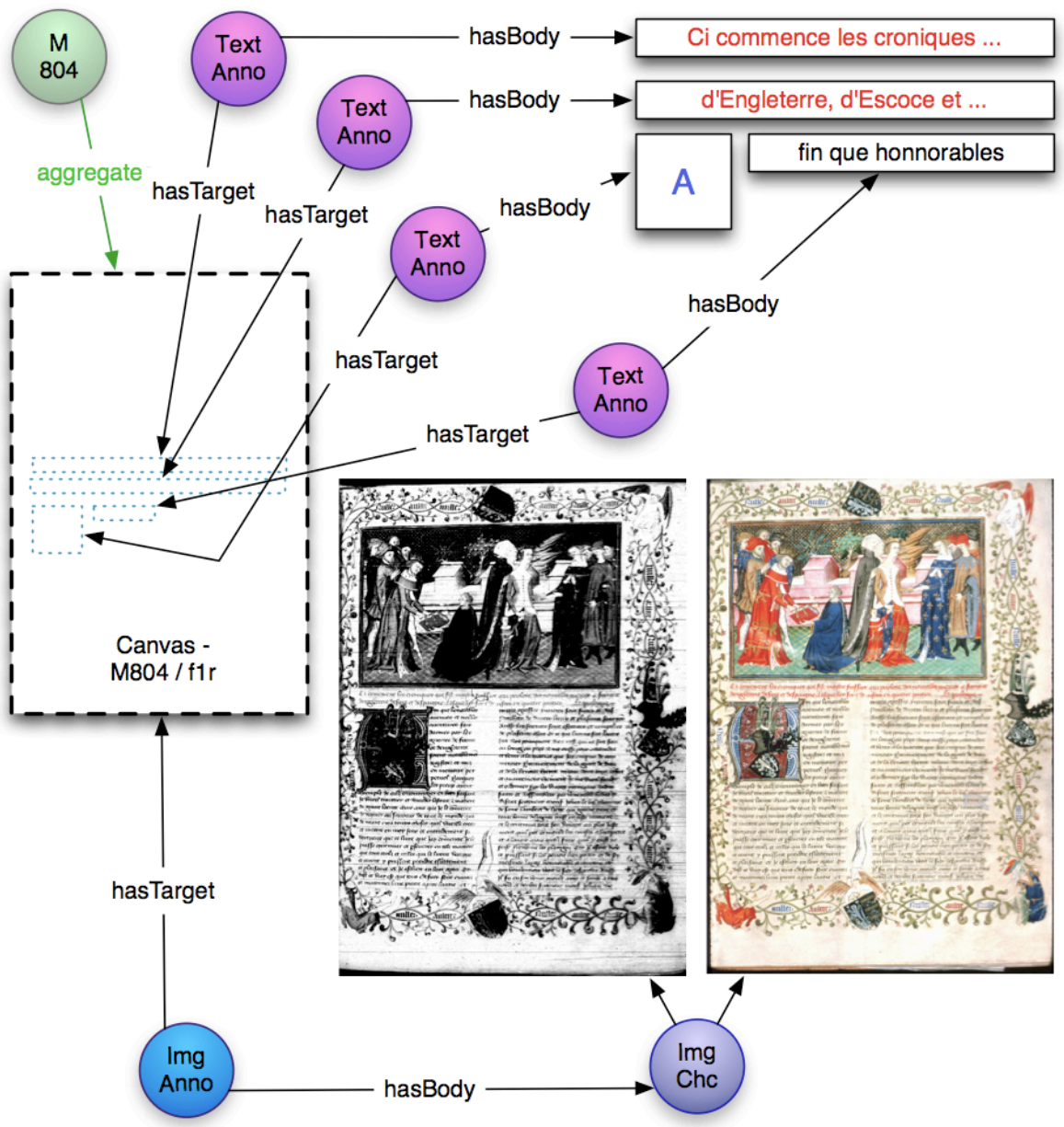

**Figure 6. Model for Morgan Library M.804, f1r [13]**

Not depicted in Figure 6 is the discovery layer, which consists of one TextOrder aggregation of annotations per page, one aggregation of the Image Annotations, and a Manifest that collects these aggregations along with the Sequence labeled "M 804". While these were not absolutely necessary, their existence greatly facilitated the creation of the experimental consuming applications, as otherwise full knowledge of the relevant triples, in the order of half a million, would be required in order to find the appropriate ones to facilitate the display of the current Canvas.

As the full text of the manuscript and scholarly notes about the illuminations and marginalia including locations are available from the previous work, this experiment provided a good demonstration that the model can be used successfully to reproduce the status quo.

### 6.1.2 *Codex Sangallensis 1394*

An extreme case of page fragments collected together comes from another page of Codex Sangallensis 1394, the manuscript depicted in Figure 2. These fragments are stuck to a white paper page, depicted along with the model in Figure 7, which is bound into the current volume. The fragments originally came from several different pages in one or more other manuscripts.

This experiment demonstrates the model's solution to non-rectangular regions on both the body (Image) and target (Canvas) of the annotations, along with the lack of a one to one relationship between images and pages, or even manuscripts. The fragments are delimited using non-rectangular regions, and Image annotations created that link to the equivalent region on a canvas.

Also shown is the mechanism for scholars to disagree about the location of fragments. The lighter image annotation, labeled "Img Ann1" and the regular "Img Ann2" both refer to the same segment of the image, but to different regions in different canvases. The display layer would be able to make a choice as to how to display this difference of opinion, again through additional metadata associated with the annotations as to author, certainty and so forth. The same approach can be applied to any other debatable annotated piece of information.

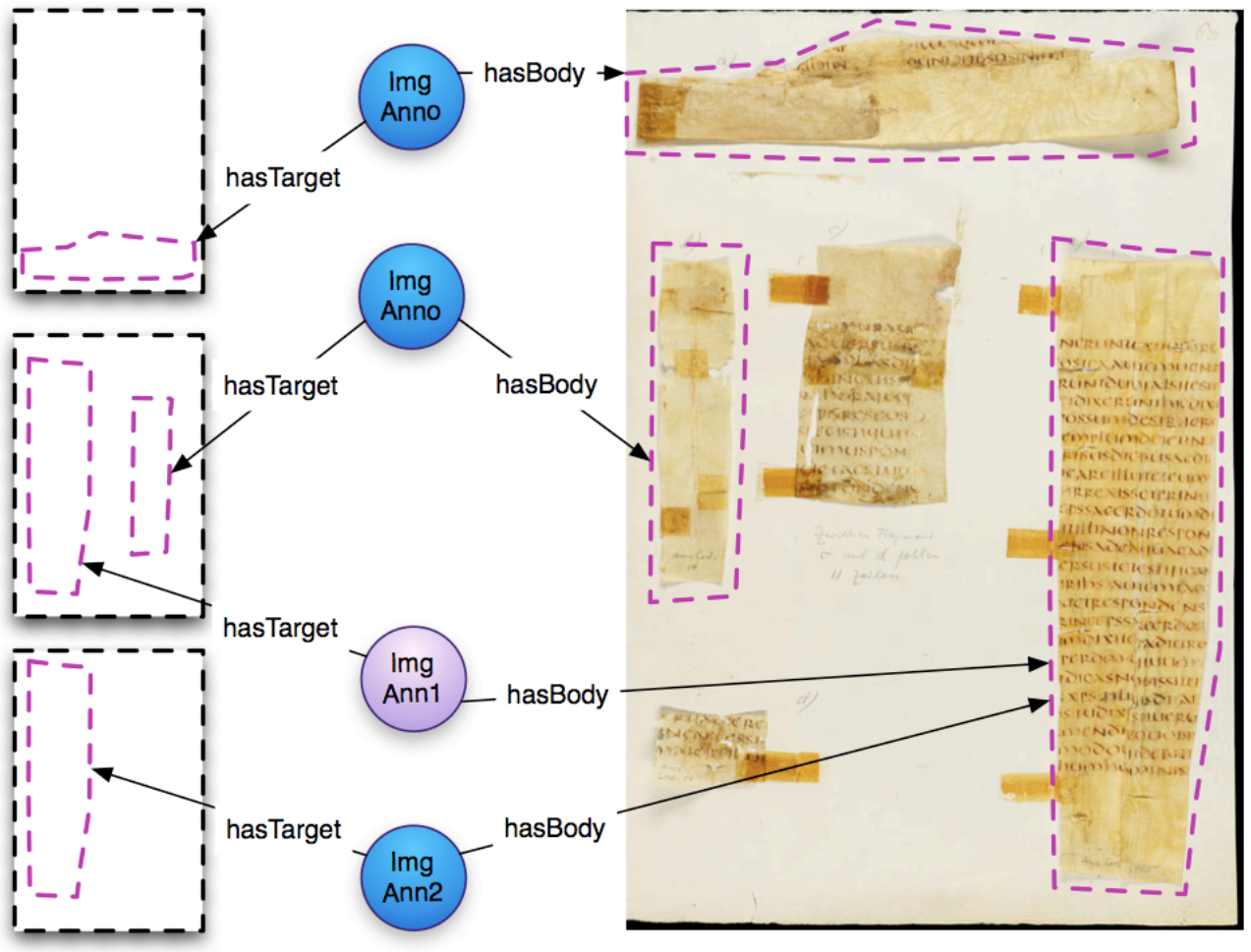

**Figure 7. Collected Fragments; Cod. Sang. 1394 p. 63 [15]**

### *6.1.3 Kantonsbibliothek Thurgau Y112*

The spread use case from Figure 1 demonstrates the need for annotations that associate Zones with multiple Canvases, and for alternate paths through the set of Canvases. The dotted lines in Figure 8 show the order in which the Canvases should be displayed: either through the spread to the top, or the two separate pages below. This order is implemented using ORE Proxies, but is too convoluted to depict.

Zone Annotations (yellow circles in the center of the diagram) are used to link any non-image annotations between the lower per-page Canvases and the appropriate locations in the spread Canvas. This ensures that all of the relevant information is maintained regardless of which path is taken without duplication.

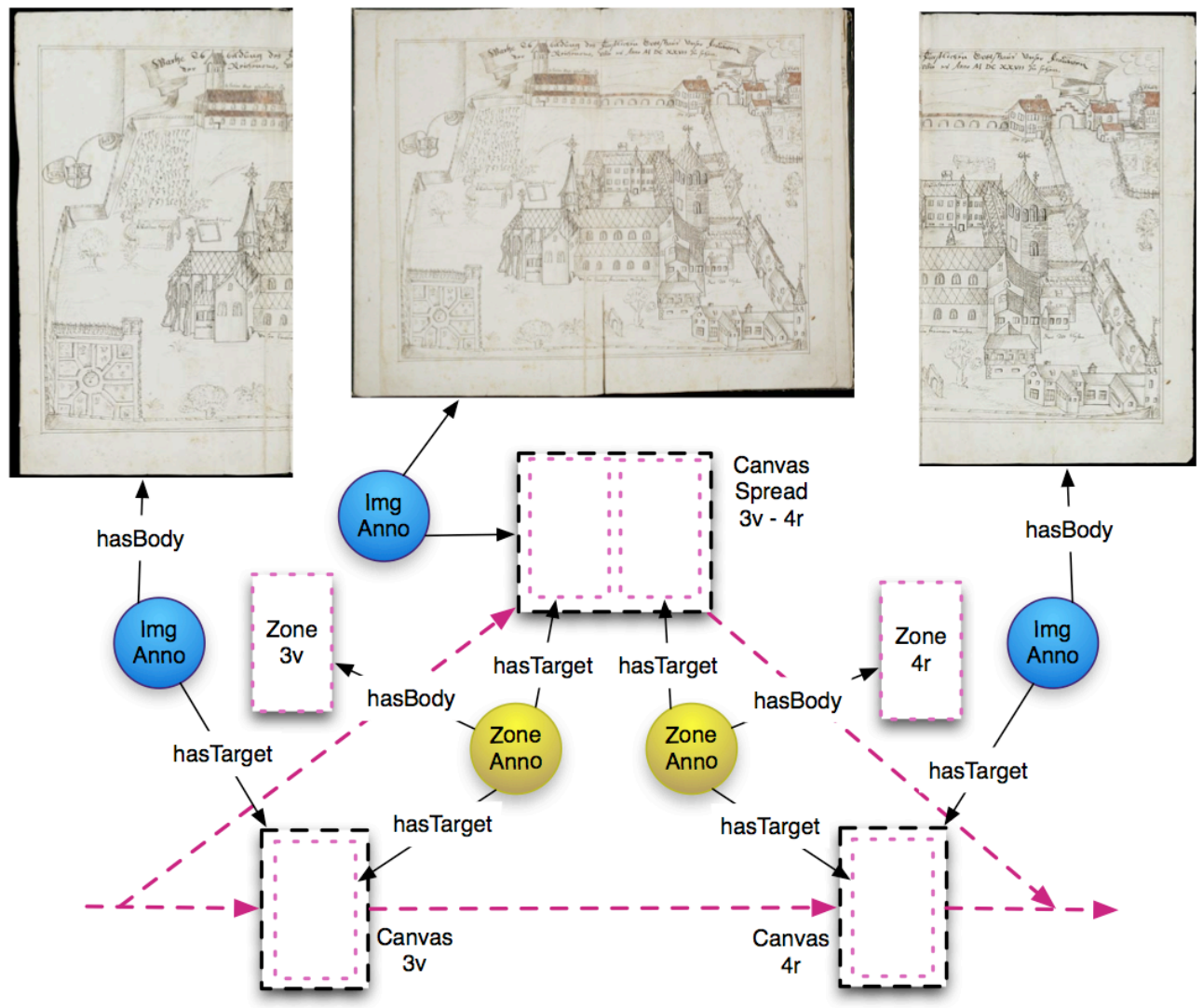

**Figure 8. Map Spread; Y112 f3v-f4r [22]**

## 6.2 Sequencing

### *6.2.1 Parker CCC 286*

Many examples exist of manuscripts that have had pages physically removed or destroyed at some point in their history. For example, Parker CCC 286 [30] is missing a leaf between the current second and third pages. To properly represent this manuscript in its original state, the fact that the page has been physically removed at some point in time must be made explicit.

Here, Canvases representing the missing leaf are added to the Sequence that describes the manuscript's original, rather than current, state. In addition, some content is known about the verso side of the missing page - it contained an illustrated frontispiece to the Book of Matthew - which can be attached to the Canvas through an annotation.

This modeling is depicted in Figure 9. There are two Sequences, "286 Curr" that represents the current state of the manuscript, and "286 Orig" that represents the original state with the additional leaf. The missing leaf has two Canvases, for recto and verso labeled Canvas 2a and 2b respectively, with Text Annotations that describe the text rather than transcribe it such as in the previous example or for the existing pages. The distinction between description and transcription would be made clear with additional properties on the annotation that are not depicted in the diagram.

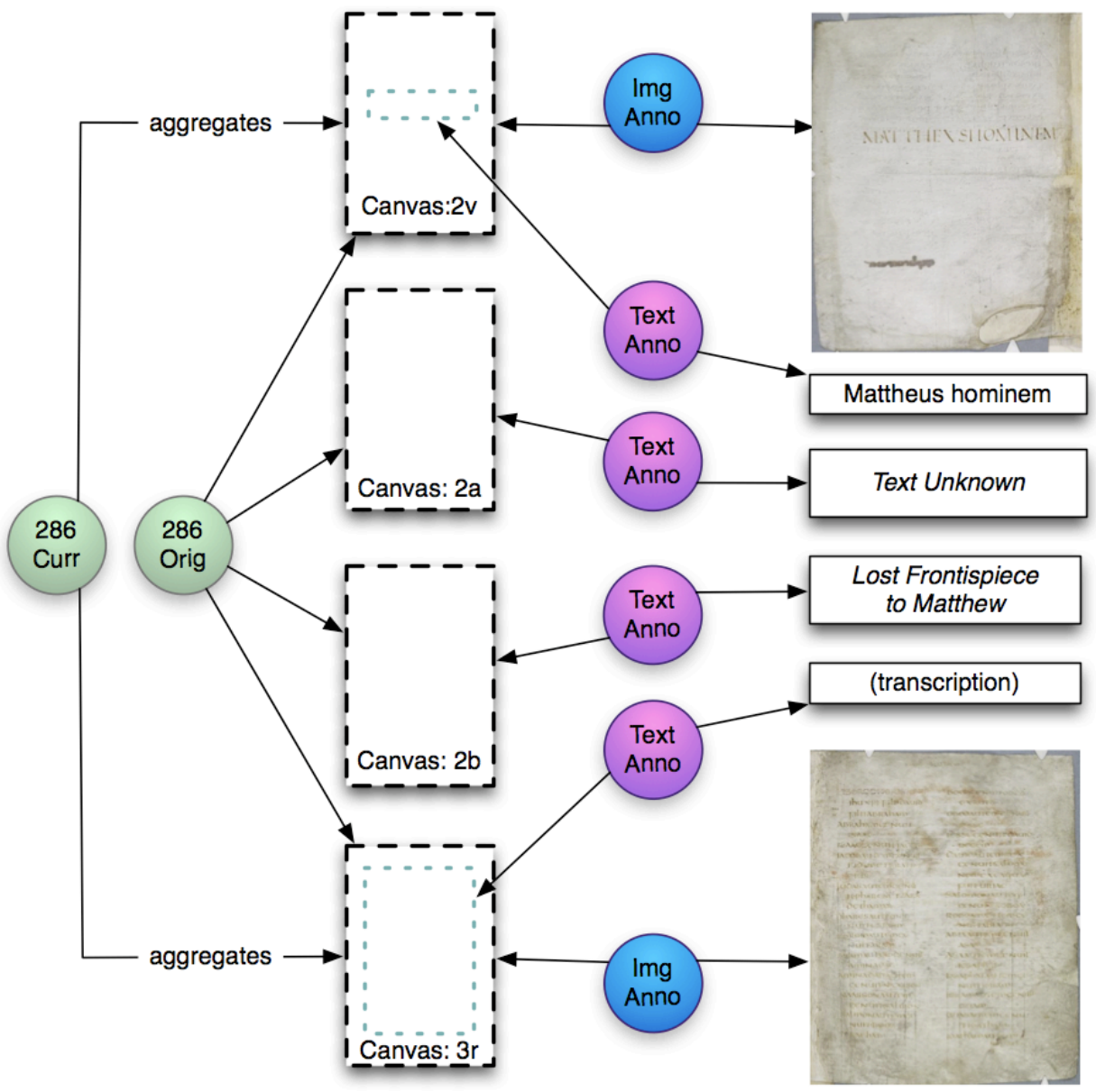

**Figure 9. Missing Pages; Parker CCC 286 [30]**

### *6.2.2 BNF F.Fr. 113-116*

An example from the Bibliothèque Nationale de France (BNF) further demonstrates the requirements for multiple Sequences over time, as well as for Ranges. The BNF currently holds four separate manuscripts, with shelfmarks Fonds Français 113 through 116, however until 1682 the text-bearing pages were bound together in a single large volume. The pages have continuous numbering from folio 1 through 735, and the text is the story of Lancelot. When the original manuscript was divided up into its current state, additional fly-leaves were added at the beginning and end of each volume, likely in order to protect the beautifully illuminated initial page.

This rebound, multi-volume text is another scenario in which multiple Sequences are crucial, but also one in which Ranges are important. Each of the individual manuscripts thus has their own Sequence (f.fr 113-116 in Figure 10), and a fifth sequence exists for the original single volume, labeled "L. du Lac". Each individual manuscript also has a Range, labeled "Content", that collects together the content bearing pages, in order to enable a display application to skip the empty pages at beginning and end. An equivalent content Range for the single volume is not needed, as the extra leaves were not present at that point.

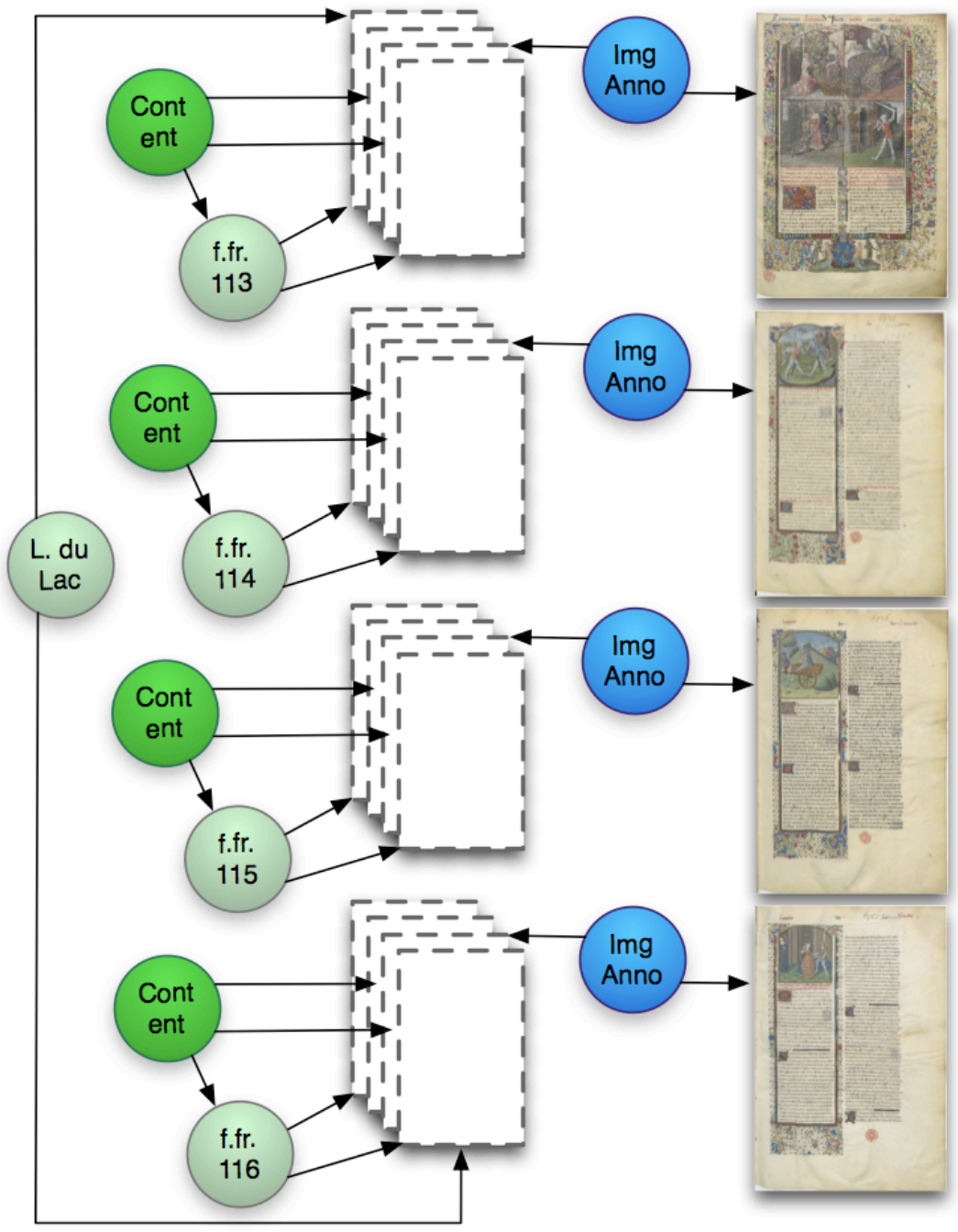

**Figure 10. Lancelot du Lac; BNF f.fr. 113-116 [21]**

## 6.3 Implementation

Implementation of the experimental descriptions was done both automatically from existing data and by hand for the more complex use cases where the existing data was not rich enough to represent what can now be expressed using the SharedCanvas model. Python's rdflib library[4] was used for production and subsequent consumption of the RDF serializations, and XSLT [10] stylesheets were explored for transforming from the internal XML formats of the Parker and e-codices collections into RDF/XML directly.

Experimental consuming applications were also implemented to demonstrate the ease of adapting existing presentation software or implementing a full suite from scratch. Three implementations were completed, including rendering in both PDF and HTML. The similarities with the PDF layout model made the transformation straightforward, with the exception of inverted canvas axes of PDF. The traditional home grown page turning application was also relatively simple to adapt, even though it originally had no concept of a Canvas. Although these implementations were not complete, they were able to duplicate and extend existing applications' functionality.

[4] http://www.rdflib.net/

# 7. CONCLUSIONS

The annotated canvas paradigm adopted in the SharedCanvas model was successful in providing solutions to the challenging use cases derived from the medieval manuscript domain. Multiple equivalent images, fragmentary pages, missing pages, different page orders over time and the description of ranges of content within a set of pages were all able to be described using only the two basic primitives of ordered ORE Aggregations and OAC Annotations.

Future work on the model will involve further experimentation with larger scale descriptions, including the Archimedes Palimpsest. It was noted during the experiments that best practice guidelines will be necessary to produce consistent models, and this will be developed as more manuscripts are described.

In the cultural heritage domain, there are many important texts recorded using media that do not have pages, such as clay tablets, or continuous scrolls. And even if it is bound in a volume today, the original object may have been a scroll that was subsequently cut up and rebound. The presentation of the reconstruction of the original form would be very different to the current paged view, even if it uses the same images. These use cases without pages can also be described with the SharedCanvas model.

In the process of designing and testing the SharedCanvas model, it was also noted that although the focus is on medieval texts, the model is reusable for any sequence of digitized images of text, regardless of era or medium. The importance of the physical object depicted in the image is obvious when it is a beautifully illuminated manuscript, however rapidly degrading 19th century newspapers such as those held at the Library of Congress[5] and British Library[6] are equally deserving of study.

In other domains, further types of data in the body of the overlay annotations would also be appropriate such as the scientific data that is depicted in a chart, or a video of a dance could be attached to the point in a text where it is described. The fact that the model could be used to describe other such resources demonstrates that it is very general at its core, and is thus likely to scale to unforeseen requirements.

Being designed from the ground up to be distributed by using the most appropriate technologies, even if they are not traditional for the domain, proved to enable the collaborative construction methods desired. Participants at the meetings from different institutions are already working together using the model to share images and the automatically extracted line segment information provided by a remote service. This distribution of information promotes a cohesive landscape of linked content, rather than each institution providing only a custom built HTML interface just for resources held in their repository.

Implementations of the model will provide unprecedented access to digital surrogates of important cultural heritage objects, distributed amongst libraries and special collections around the world. By using the web to bring the humanists' primary data and the related scholarship to them, their research capacity is enhanced and the shared annotation space brings them together with other scholars working in the same realm.

[5] http://chroniclingamerica.loc.gov/ and http://www.loc.gov/ndnp/

[6] http://newspapers.bl.uk/blcs/

# ACKNOWLEDGMENTS

This research was funded by the Andrew W. Mellon Foundation, through the OAC and Stanford's Digitized Medieval Manuscript grants. The authors would like to thank the participants at the DMSTech meetings, and the institutions that own the manuscripts for allowing their use in this publication, including Christopher De Hamel, Donnelly Fellow Librarian, and the Master and Fellows of Corpus Christi College, Cambridge.